\documentclass[aps,prl,twocolumn,superscriptaddress]{revtex4-2}  % for review and submission
\usepackage{graphicx,comment,float,braket,textcomp,amsmath,lipsum,color,soul}  % needed for figures
\usepackage{dcolumn}   % needed for some tables
\usepackage{bm}        % for math
\usepackage{amssymb}   % for math

\begin{document}

%\title{Wavefunction Engineering of Nonlocal Optical Nonlinearities}

% is using macroscopic ok in this situation? Seems as micrometer length scales count as macroscopic:
% Lau and Clerk - "Macroscale entanglement and measurement"
\title{Nonlocal Coherent Optical Nonlinearities of a Macroscopic Quantum System}

\author{Albert Liu}
\affiliation{Condensed Matter Physics and Materials Science Division, Brookhaven National Laboratory, Upton, New York, 11973 USA}

\author{Eric W. Martin}
\thanks{Current Affiliation: MONSTR Sense Technologies, LLC, Ann Arbor, Michigan 48108, USA}
\affiliation{Department of Physics, University of Michigan, Ann Arbor, Michigan 48109, United States}

\author{Jiaqi Hu}
\affiliation{Department of Physics, University of Michigan, Ann Arbor, Michigan 48109, United States}

\author{Zhaorong Wang}
\affiliation{Department of Physics, University of Michigan, Ann Arbor, Michigan 48109, United States}

\author{Hui Deng}
\affiliation{Department of Physics, University of Michigan, Ann Arbor, Michigan 48109, United States}
\affiliation{Quantum Research Institute, University of Michigan, Ann Arbor, Michigan 48109, United States}

\author{Steven T. Cundiff}
\email{cundiff@umich.edu} 
\affiliation{Department of Physics, University of Michigan, Ann Arbor, Michigan 48109, United States}
\affiliation{Quantum Research Institute, University of Michigan, Ann Arbor, Michigan 48109, United States}

\vskip 0.3cm

\date{\today}

\begin{abstract}
The optical responses of solids are typically understood to be local in space. Whether locality holds for the optical response of a macroscopic quantum system has, however, been largely unexplored. Here, we use multidimensional coherent spectroscopy at the optical diffraction limit to demonstrate nonlocal optical nonlinearities in a semiconductor microcavity. These nonlocal optical responses are both coherent and quantum in nature, deriving from the macroscopic length scale of confined exciton-polariton wavefunctions.
\end{abstract}

\maketitle

In standard treatments of light-matter interactions, the material response is assumed to be local in space. Despite its ubiquity, the limitations of this assumption are of both fundamental and practical importance. For example, it is well-known that optical responses can exhibit non-locality when length scales reach the near-field regime \cite{Henkel2006,Mortensen2014,Eriksen2024}. Nonlocal optical responses can also occur when mobile excitations travel to other points in space for example via photorefraction \cite{Segev1992}, thermal \cite{Rotschild2005} and charge diffusion \cite{Pizzuto2021}, or phonon-polariton propagation \cite{Henstridge2022}. To manifest any of the above effects that rely on energy or charge transport, however, the optical medium need not be quantum (i.e. no macroscopic quantum coherence is necessary).

Theoretical predictions for specially-prepared atom-waveguide systems \cite{Shahmoon2016} have pointed to quantum effects as a potential new source of nonlocal optical responses. Yet intuitively, any macroscopic quantum state should naturally exhibit nonlocal optical responses by virtue of a delocalized wavefunction. 
In real materials, however, short coherence lengths typically preclude observation of such nonlocality. Exciton-polaritons, hybrid excitations between electron-hole pairs and light \cite{Basov2021}, are an exception to this conventional wisdom. With coherence lengths reaching macroscopic length scales, spectacular demonstrations of quantum effects have been reported, ranging from Josephson oscillations \cite{Lagoudakis2010,Abbarchi2013} to Bose-Einstein condensation \cite{Deng2002,Kasprzak2006}.

\begin{figure*}[t]
    \centering
    \includegraphics[width=0.97\linewidth]{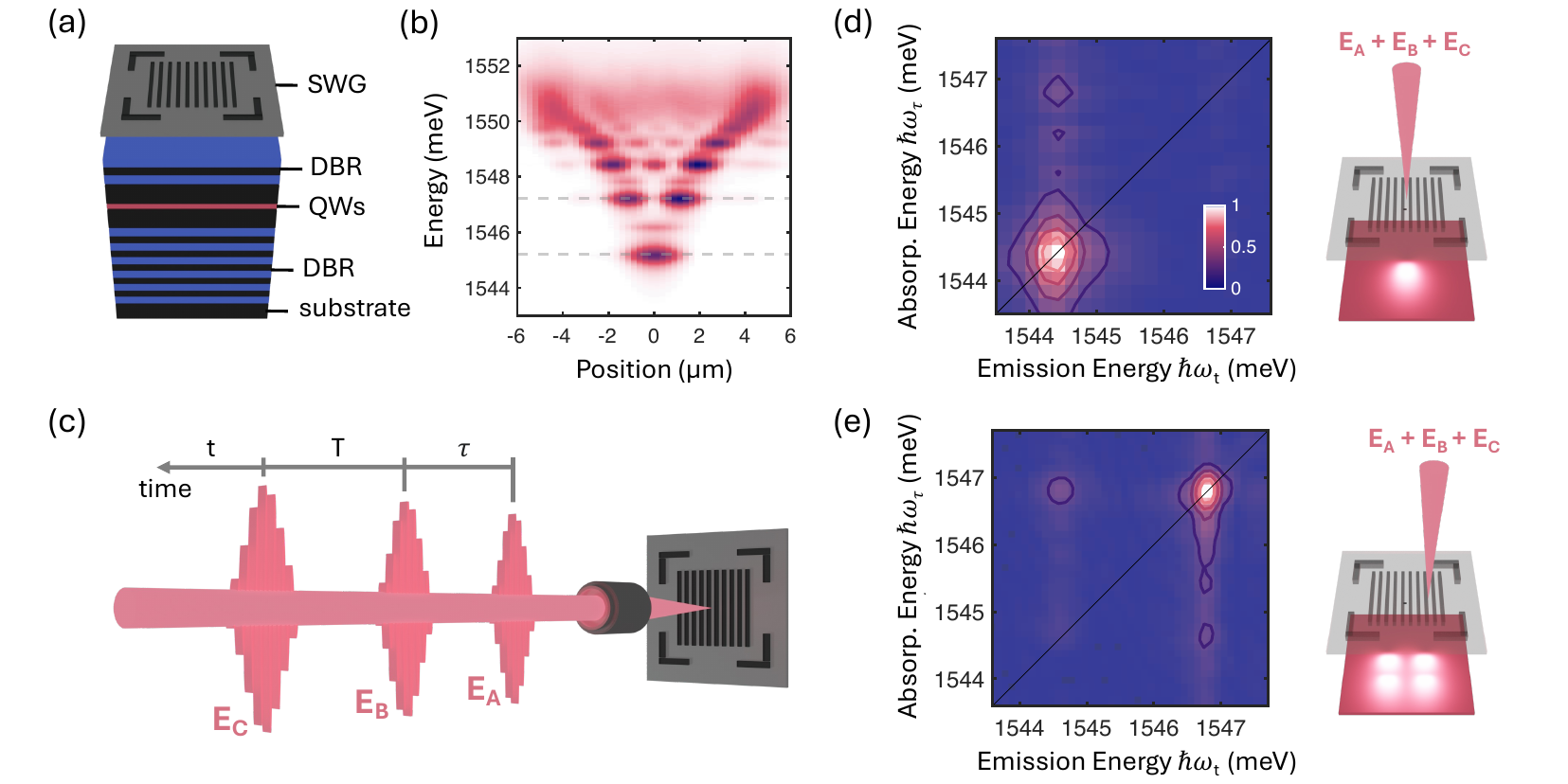}
    \caption{(a) Schematic of the hybrid cavity comprised of quantum wells (QWs) between distributed Bragg reflectors (DBRs) and a sub-wavelength grating (SWG) mirror. (b) Spectrally-resolved real-space image of photoluminescence from the device in (a), showing the spatial profiles of the confined lower-branch polariton modes. Dashed lines marked the first and second excited-states probed here. (c) Schematic of the MDCS measurement, in which three excitation pulses $\{E_A,E_B,E_C\}$ are focused by a microscope objective ($NA = 0.5$) on the sample. The nonlinear signal is measured as a function of time-delays $\{\tau,t\}$ with $T = 200$ fs for all measurements. (d,e) Single-quantum spectra with the excitation beams positioned at the (d) center of the SWG (antinode of the first excited-state) and (e) 1.5 $\mu$m to the right and 0.5 $\mu$m below (antinode of the second excited-state). Cartoons depict the excitation beams targeting specific exciton-polariton wavefunctions in the quantum wells (red squares).}
    \label{Fig1}
\end{figure*}

As composite light-matter particles, polaritons are also amenable to photonic engineering techniques. For example, exciton-polaritons in a semiconductor microcavity can be laterally-confined by sharp discontinuities in the cavity refractive index. An elegant way this has been accomplished is by use of a sub-wavelength grating (SWG) as a cavity mirror, in which the grating dimensions impose an effective confinement potential \cite{Zhang2014,Zhang2015}. The resultant micrometer length-scale exciton-polariton wavefunctions are a perfect host for {\it nonlocal} nonlinear optics that rely on macroscopic quantum coherence rather than transport effects, which we demonstrate in this Letter.

The system studied is illustrated in Fig.~\ref{Fig1}(a), comprised of GaAs quantum wells (QWs) embedded in a cavity formed by Al$_{0.15}$GaAs/AlAs distributed Bragg reflector (DBR) layers and a square SWG that acts as a high-reflector for light polarized along the grating bars (referred to here as TE-polarization). The TE-polarized photoluminescence reveals three-dimensional quantum confinement of the exciton-polaritons, as shown by the spectrally-resolved real-space image in Fig.~\ref{Fig1}(b). The lower polariton branch is split into discrete eigenstates due to the lateral confinement, whose energies and spatial distributions are reproduced by a harmonic confinement potential \cite{Zhang2015}. The macroscopic polariton states observed (with spatial extent exceeding the corresponding optical diffraction limit) can now be selectively excited by positioning optical excitation at the antinodes of their wavefunction. For simplicity, here we restrict our attention to the two lowest-energy states indicated by the dashed lines in Fig.~\ref{Fig1}(b).

While the linear optical properties of these zero-dimensional polaritons are now well-understood, their nonlinear optical properties remain unexplored. Here we apply the technique of multidimensional coherent spectroscopy (MDCS), capable of characterizing the full multidimensional nonlinear optical response of a given system \cite{MDCS_Book}. MDCS has been used to study many aspects of polaritonic physics, ranging from polariton coupling \cite{Takemura2015,Xiang2024} to their higher-order excitations \cite{Wen2013,Autry2020}, but its application to quantum-confined polaritonic systems has not yet been reported. 

A schematic of our MDCS experiment is shown in Fig.~\ref{Fig1}(c) in which three  TE-polarized excitation pulses $\{E_A,E_B,E_C\}$ (with sufficient bandwidth to excite all of the states identified by photoluminescence) impinge on the sample at normal incidence, cooperatively generating a coherent electric field as a function of the time delays $\{\tau,T,t\}$ as indicated. We note that the collinear excitation geometry used here \cite{Nardin2013,Takemura2015,Martin2020} (in contrast to the more common non-collinear geometry \cite{MDCS_Book}) is crucial to reaching a near diffraction-limited excitation spot size ($\approx$ 1$\mu$m). By measuring the nonlinear electric field along two time delays and performing a two-dimensional (2-D) Fourier transform, the dynamics along each time delay are correlated in a 2-D spectrum. Various types of 2-D spectra are possible \cite{MDCS_Book} that reveal unique aspects of a system's microscopic physics, but here we focus on the spectrum obtained by Fourier transforming along the delays $\{\tau,t\}$. These single-quantum spectra correlate the photon energies of initial optical absorption and subsequent optical emission in a nonlinear wave-mixing process, which are ideal for measuring both coherent and incoherent coupling between optical transitions. Note that, due to minimal inhomogeneous broadening, the two types of single-quantum spectra (rephasing and non-rephasing) provide identical information (see Supplemental Material for comparison). We therefore consider non-rephasing spectra in the following, due to a slight advantage in peak clarity.

\begin{figure}[h]
    \centering
    \includegraphics[width=0.95\linewidth]{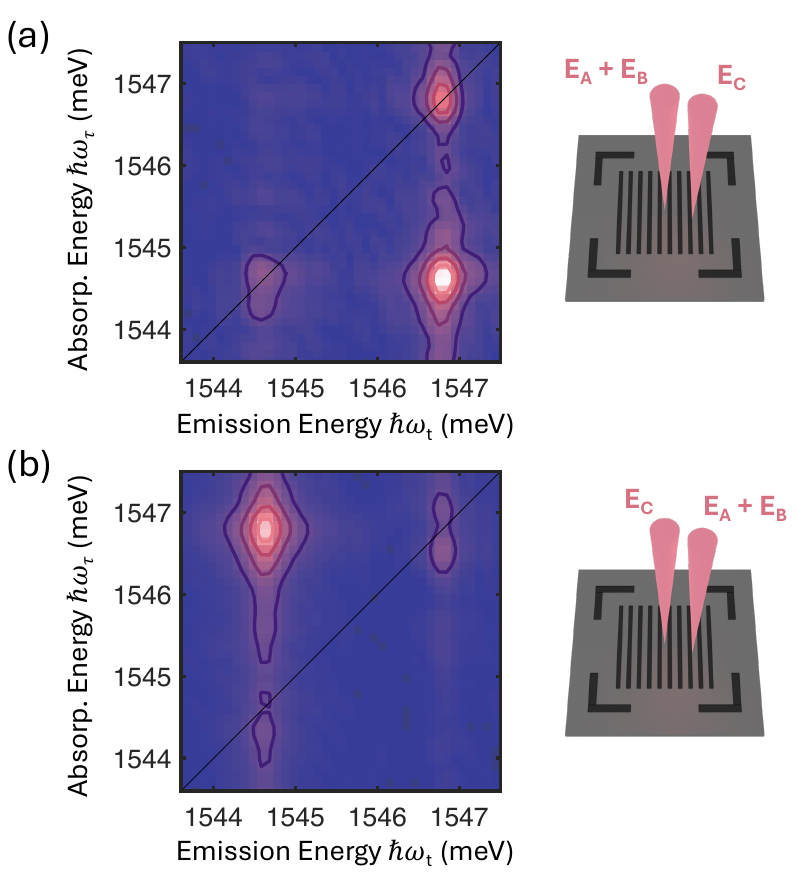}
    \caption{(a,b) Single-quantum spectra acquired with the `pump' ($E_A+E_B$) and `probe' ($E_C$) pulses exciting different spatial locations on the sample. (a) The (probe)pump pulses interact with the (second-)first-excited state. (b) Spatial locations of the pump and probe pulses are exchanged.}
    \label{Fig2}
\end{figure}

To demonstrate the spatial selectivity of confined exciton-polariton wavefunctions, single-quantum spectra acquired with excitation at two distinct sample positions are plotted in Figs.~\ref{Fig1}(d) and \ref{Fig1}(e). In the spectrum shown in Fig.~\ref{Fig1}(d), all excitation beams are positioned at the anti-node of the first excited-state mode (center of the SWG). A single peak is observed that arises from optical absorption (vertical axis $\hbar\omega_\tau$) and optical emission (horizontal axis $\hbar\omega_t$) at the first excited-state polariton energy. In the spectrum shown in Fig.~\ref{Fig1}(e), the excitation beams are translated (1.5$\mu$m laterally and 0.5$\mu$m vertically away from the center position) to an anti-node of the second excited-state mode. A single dominant peak is again observed, this time arising from optical absorption and optical emission at the second excited-state polariton energy.

Having demonstrated that each exciton-polariton wavefunction offers a locally-distinct nonlinear optical response, we now investigate the possibility of a nonlocal nonlinear optical response {\it between} different wavefunctions. Referring to the first two excitation pulses ($E_A + E_B$) as the `pump' and the third excitation pulse ($E_C$) as the `probe', spectra acquired with the pump and probe pulses separated in space are plotted in Fig.~\ref{Fig2}. For the spectrum in Fig.~\ref{Fig2}(a), the pump pulses are spatially centered at the anti-node of the first excited-state while the probe pulse is centered at that of the second excited-state. With this excitation scheme a new peak now emerges whose position corresponds to absorption at the first excited-state polariton energy and emission at the second excited-state polariton energy, {\it direct evidence of nonlocal optical nonlinearities between two spatial quantum wavefunctions}. By exchanging pump and probe positions, the inverse process of absorption at the second excited-state energy and emission at the first excited-state energy is observed in the spectrum in Fig.~\ref{Fig2}(b). These observations demonstrate that the coupling we observe via MDCS is not simply energy relaxation between the two states, but is indeed true coherent coupling in which optical absorption by one resonance modifies the optical response of the other at a different spatial location.

\begin{figure}[b]
    \centering
    \includegraphics[width=0.95\linewidth]{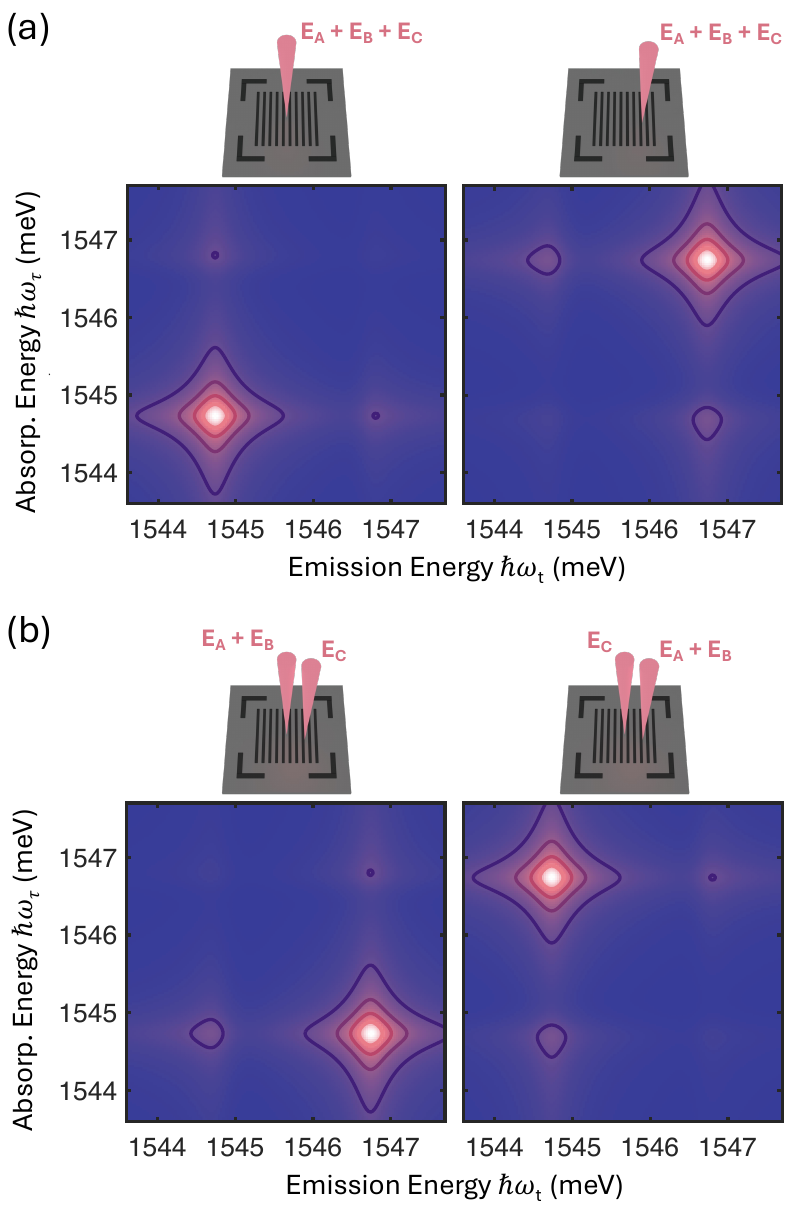}
    \caption{Perturbative density matrix simulations of (a) local and (b) nonlocal exciton-polariton nonlinearities. Identical local dipole moments, estimated from the real-space photoluminescence intensity in Fig.~\ref{Fig1}(b), and an assumed dephasing rate of 0.3 ps$^{-1}$ are used for all simulations.}
    \label{Fig3}
\end{figure}

To confirm that the observed local and nonlocal nonlinearities indeed derive from the spatial-dependence of distinct exciton-polariton wavefunctions, we perform simulations of the single-quantum MDCS spectra based on a standard perturbative solution of the system density matrix dynamics \cite{MDCS_Book}. While full electrodynamic simulations of each two-dimensional exciton-polariton state was prohibitively difficult, the spatial-dependences of their transition dipole moments can be well-estimated from their respective photoluminescence intensity profiles in Fig.~\ref{Fig1}(b). For each exciton-polariton state, the square root of their intensity profile is used as a proxy of the local transition dipole moment and is assumed to be symmetric along and across the grating bars. We further account for the finite spatial extent of optical excitation by averaging each local dipole moment over a 1 $\mu$m spot size, and assume a 0.3 ps$^{-1}$ dephasing rate for both states considered. The resultant simulations are plotted in Fig.~\ref{Fig3}, which agree well with the corresponding experimental spectra shown in Figs.~\ref{Fig1} and \ref{Fig2}. We note that population relaxation and higher-lying states are neglected in the simulations, but do not contribute to the salient physics considered here.

In conclusion, we demonstrate nonlocal optical nonlinearities between two distinct exciton-polariton wavefunctions. This demonstration of a nonlocal coherent optical response in a macroscopic quantum system, raises possibilities of nonlocal optical nonlinearities in other quantum material platforms, for example in superconductors with terahertz optical responses \cite{Liu_2024_NatPhys}. Here we considered third-order nonlinearities between two polaritonic states, but the possibilities expand dramatically by considering higher-order nonlinearities and higher-energy wavefunctions. With further refinement of the technique presented here, we envisage the generation of designer spatial emission patterns using these nonlocal nonlinearities and a new paradigm of coherent {\it spatial} control. 

{\it Acknowledgments} - The work at University of Michigan was supported by the National Science Foundation (NSF)
under Grant No. 1622768. Albert Liu was supported by the US Department of Energy (DOE), Office of Basic Energy Sciences (BES) under contract no. DOE-SC0012704.

{\it Disclosures} - EWM: MONSTr Sense Technologies LLC (I,E), STC: MONSTr Sense Technologies LLC (I,E).

\bibliography{bibliography}

%apsrev4-2.bst 2019-01-14 (MD) hand-edited version of apsrev4-1.bst
%Control: key (0)
%Control: author (8) initials jnrlst
%Control: editor formatted (1) identically to author
%Control: production of article title (0) allowed
%Control: page (0) single
%Control: year (1) truncated
%Control: production of eprint (0) enabled
\begin{thebibliography}{23}%
\makeatletter
\providecommand \@ifxundefined [1]{%
 \@ifx{#1\undefined}
}%
\providecommand \@ifnum [1]{%
 \ifnum #1\expandafter \@firstoftwo
 \else \expandafter \@secondoftwo
 \fi
}%
\providecommand \@ifx [1]{%
 \ifx #1\expandafter \@firstoftwo
 \else \expandafter \@secondoftwo
 \fi
}%
\providecommand \natexlab [1]{#1}%
\providecommand \enquote  [1]{``#1''}%
\providecommand \bibnamefont  [1]{#1}%
\providecommand \bibfnamefont [1]{#1}%
\providecommand \citenamefont [1]{#1}%
\providecommand \href@noop [0]{\@secondoftwo}%
\providecommand \href [0]{\begingroup \@sanitize@url \@href}%
\providecommand \@href[1]{\@@startlink{#1}\@@href}%
\providecommand \@@href[1]{\endgroup#1\@@endlink}%
\providecommand \@sanitize@url [0]{\catcode `\\12\catcode `\$12\catcode
  `\&12\catcode `\#12\catcode `\^12\catcode `\_12\catcode `\%12\relax}%
\providecommand \@@startlink[1]{}%
\providecommand \@@endlink[0]{}%
\providecommand \url  [0]{\begingroup\@sanitize@url \@url }%
\providecommand \@url [1]{\endgroup\@href {#1}{\urlprefix }}%
\providecommand \urlprefix  [0]{URL }%
\providecommand \Eprint [0]{\href }%
\providecommand \doibase [0]{https://doi.org/}%
\providecommand \selectlanguage [0]{\@gobble}%
\providecommand \bibinfo  [0]{\@secondoftwo}%
\providecommand \bibfield  [0]{\@secondoftwo}%
\providecommand \translation [1]{[#1]}%
\providecommand \BibitemOpen [0]{}%
\providecommand \bibitemStop [0]{}%
\providecommand \bibitemNoStop [0]{.\EOS\space}%
\providecommand \EOS [0]{\spacefactor3000\relax}%
\providecommand \BibitemShut  [1]{\csname bibitem#1\endcsname}%
\let\auto@bib@innerbib\@empty
%</preamble>
\bibitem [{\citenamefont {Henkel}\ and\ \citenamefont
  {Joulain}(2006)}]{Henkel2006}%
  \BibitemOpen
  \bibfield  {author} {\bibinfo {author} {\bibfnamefont {C.}~\bibnamefont
  {Henkel}}\ and\ \bibinfo {author} {\bibfnamefont {K.}~\bibnamefont
  {Joulain}},\ }\bibfield  {title} {\bibinfo {title} {Electromagnetic field
  correlations near a surface with a nonlocal optical response},\ }\href
  {https://doi.org/10.1007/s00340-006-2219-9} {\bibfield  {journal} {\bibinfo
  {journal} {Appl. Phys. B}\ }\textbf {\bibinfo {volume} {84}},\ \bibinfo
  {pages} {61} (\bibinfo {year} {2006})}\BibitemShut {NoStop}%
\bibitem [{\citenamefont {Mortensen}\ \emph {et~al.}(2014)\citenamefont
  {Mortensen}, \citenamefont {Raza}, \citenamefont {Wubs}, \citenamefont
  {S{\o}ndergaard},\ and\ \citenamefont {Bozhevolnyi}}]{Mortensen2014}%
  \BibitemOpen
  \bibfield  {author} {\bibinfo {author} {\bibfnamefont {N.~A.}\ \bibnamefont
  {Mortensen}}, \bibinfo {author} {\bibfnamefont {S.}~\bibnamefont {Raza}},
  \bibinfo {author} {\bibfnamefont {M.}~\bibnamefont {Wubs}}, \bibinfo {author}
  {\bibfnamefont {T.}~\bibnamefont {S{\o}ndergaard}},\ and\ \bibinfo {author}
  {\bibfnamefont {S.~I.}\ \bibnamefont {Bozhevolnyi}},\ }\bibfield  {title}
  {\bibinfo {title} {A generalized non-local optical response theory for
  plasmonic nanostructures},\ }\href {https://doi.org/10.1038/ncomms4809}
  {\bibfield  {journal} {\bibinfo  {journal} {Nat. Commun.}\ }\textbf {\bibinfo
  {volume} {5}},\ \bibinfo {pages} {3809} (\bibinfo {year} {2014})}\BibitemShut
  {NoStop}%
\bibitem [{\citenamefont {Eriksen}\ \emph {et~al.}(2024)\citenamefont
  {Eriksen}, \citenamefont {Tserkezis}, \citenamefont {Mortensen},\ and\
  \citenamefont {Cox}}]{Eriksen2024}%
  \BibitemOpen
  \bibfield  {author} {\bibinfo {author} {\bibfnamefont {M.~H.}\ \bibnamefont
  {Eriksen}}, \bibinfo {author} {\bibfnamefont {C.}~\bibnamefont {Tserkezis}},
  \bibinfo {author} {\bibfnamefont {N.~A.}\ \bibnamefont {Mortensen}},\ and\
  \bibinfo {author} {\bibfnamefont {J.~D.}\ \bibnamefont {Cox}},\ }\bibfield
  {title} {\bibinfo {title} {Nonlocal effects in plasmon-emitter
  interactions},\ }\href {https://doi.org/doi:10.1515/nanoph-2023-0575}
  {\bibfield  {journal} {\bibinfo  {journal} {Nanophotonics}\ }\textbf
  {\bibinfo {volume} {13}},\ \bibinfo {pages} {2741} (\bibinfo {year}
  {2024})}\BibitemShut {NoStop}%
\bibitem [{\citenamefont {Segev}\ \emph {et~al.}(1992)\citenamefont {Segev},
  \citenamefont {Crosignani}, \citenamefont {Yariv},\ and\ \citenamefont
  {Fischer}}]{Segev1992}%
  \BibitemOpen
  \bibfield  {author} {\bibinfo {author} {\bibfnamefont {M.}~\bibnamefont
  {Segev}}, \bibinfo {author} {\bibfnamefont {B.}~\bibnamefont {Crosignani}},
  \bibinfo {author} {\bibfnamefont {A.}~\bibnamefont {Yariv}},\ and\ \bibinfo
  {author} {\bibfnamefont {B.}~\bibnamefont {Fischer}},\ }\bibfield  {title}
  {\bibinfo {title} {Spatial solitons in photorefractive media},\ }\href
  {https://doi.org/10.1103/PhysRevLett.68.923} {\bibfield  {journal} {\bibinfo
  {journal} {Phys. Rev. Lett.}\ }\textbf {\bibinfo {volume} {68}},\ \bibinfo
  {pages} {923} (\bibinfo {year} {1992})}\BibitemShut {NoStop}%
\bibitem [{\citenamefont {Rotschild}\ \emph {et~al.}(2005)\citenamefont
  {Rotschild}, \citenamefont {Cohen}, \citenamefont {Manela}, \citenamefont
  {Segev},\ and\ \citenamefont {Carmon}}]{Rotschild2005}%
  \BibitemOpen
  \bibfield  {author} {\bibinfo {author} {\bibfnamefont {C.}~\bibnamefont
  {Rotschild}}, \bibinfo {author} {\bibfnamefont {O.}~\bibnamefont {Cohen}},
  \bibinfo {author} {\bibfnamefont {O.}~\bibnamefont {Manela}}, \bibinfo
  {author} {\bibfnamefont {M.}~\bibnamefont {Segev}},\ and\ \bibinfo {author}
  {\bibfnamefont {T.}~\bibnamefont {Carmon}},\ }\bibfield  {title} {\bibinfo
  {title} {Solitons in nonlinear media with an infinite range of nonlocality:
  First observation of coherent elliptic solitons and of vortex-ring
  solitons},\ }\href {https://doi.org/10.1103/PhysRevLett.95.213904} {\bibfield
   {journal} {\bibinfo  {journal} {Phys. Rev. Lett.}\ }\textbf {\bibinfo
  {volume} {95}},\ \bibinfo {pages} {213904} (\bibinfo {year}
  {2005})}\BibitemShut {NoStop}%
\bibitem [{\citenamefont {Pizzuto}\ \emph {et~al.}(2021)\citenamefont
  {Pizzuto}, \citenamefont {Castro-Camus}, \citenamefont {Wilson},
  \citenamefont {Choi}, \citenamefont {Li},\ and\ \citenamefont
  {Mittleman}}]{Pizzuto2021}%
  \BibitemOpen
  \bibfield  {author} {\bibinfo {author} {\bibfnamefont {A.}~\bibnamefont
  {Pizzuto}}, \bibinfo {author} {\bibfnamefont {E.}~\bibnamefont
  {Castro-Camus}}, \bibinfo {author} {\bibfnamefont {W.}~\bibnamefont
  {Wilson}}, \bibinfo {author} {\bibfnamefont {W.}~\bibnamefont {Choi}},
  \bibinfo {author} {\bibfnamefont {X.}~\bibnamefont {Li}},\ and\ \bibinfo
  {author} {\bibfnamefont {D.~M.}\ \bibnamefont {Mittleman}},\ }\bibfield
  {title} {\bibinfo {title} {Nonlocal time-resolved terahertz spectroscopy in
  the near field},\ }\href {https://doi.org/10.1021/acsphotonics.1c01367}
  {\bibfield  {journal} {\bibinfo  {journal} {ACS Photon.}\ }\textbf {\bibinfo
  {volume} {8}},\ \bibinfo {pages} {2904} (\bibinfo {year} {2021})}\BibitemShut
  {NoStop}%
\bibitem [{\citenamefont {Henstridge}\ \emph {et~al.}(2022)\citenamefont
  {Henstridge}, \citenamefont {F{\"o}rst}, \citenamefont {Rowe}, \citenamefont
  {Fechner},\ and\ \citenamefont {Cavalleri}}]{Henstridge2022}%
  \BibitemOpen
  \bibfield  {author} {\bibinfo {author} {\bibfnamefont {M.}~\bibnamefont
  {Henstridge}}, \bibinfo {author} {\bibfnamefont {M.}~\bibnamefont
  {F{\"o}rst}}, \bibinfo {author} {\bibfnamefont {E.}~\bibnamefont {Rowe}},
  \bibinfo {author} {\bibfnamefont {M.}~\bibnamefont {Fechner}},\ and\ \bibinfo
  {author} {\bibfnamefont {A.}~\bibnamefont {Cavalleri}},\ }\bibfield  {title}
  {\bibinfo {title} {Nonlocal nonlinear phononics},\ }\href
  {https://doi.org/10.1038/s41567-022-01512-3} {\bibfield  {journal} {\bibinfo
  {journal} {Nat. Phys.}\ }\textbf {\bibinfo {volume} {18}},\ \bibinfo {pages}
  {457} (\bibinfo {year} {2022})}\BibitemShut {NoStop}%
\bibitem [{\citenamefont {Shahmoon}\ \emph {et~al.}(2016)\citenamefont
  {Shahmoon}, \citenamefont {Gri\v{s}ins}, \citenamefont {Stimming},
  \citenamefont {Mazets},\ and\ \citenamefont {Kurizki}}]{Shahmoon2016}%
  \BibitemOpen
  \bibfield  {author} {\bibinfo {author} {\bibfnamefont {E.}~\bibnamefont
  {Shahmoon}}, \bibinfo {author} {\bibfnamefont {P.}~\bibnamefont
  {Gri\v{s}ins}}, \bibinfo {author} {\bibfnamefont {H.~P.}\ \bibnamefont
  {Stimming}}, \bibinfo {author} {\bibfnamefont {I.}~\bibnamefont {Mazets}},\
  and\ \bibinfo {author} {\bibfnamefont {G.}~\bibnamefont {Kurizki}},\
  }\bibfield  {title} {\bibinfo {title} {Highly nonlocal optical nonlinearities
  in atoms trapped near a waveguide},\ }\href
  {https://doi.org/10.1364/OPTICA.3.000725} {\bibfield  {journal} {\bibinfo
  {journal} {Optica}\ }\textbf {\bibinfo {volume} {3}},\ \bibinfo {pages} {725}
  (\bibinfo {year} {2016})}\BibitemShut {NoStop}%
\bibitem [{\citenamefont {Basov}\ \emph {et~al.}(2021)\citenamefont {Basov},
  \citenamefont {Asenjo-Garcia}, \citenamefont {Schuck}, \citenamefont {Zhu},\
  and\ \citenamefont {Rubio}}]{Basov2021}%
  \BibitemOpen
  \bibfield  {author} {\bibinfo {author} {\bibfnamefont {D.~N.}\ \bibnamefont
  {Basov}}, \bibinfo {author} {\bibfnamefont {A.}~\bibnamefont
  {Asenjo-Garcia}}, \bibinfo {author} {\bibfnamefont {P.~J.}\ \bibnamefont
  {Schuck}}, \bibinfo {author} {\bibfnamefont {X.}~\bibnamefont {Zhu}},\ and\
  \bibinfo {author} {\bibfnamefont {A.}~\bibnamefont {Rubio}},\ }\bibfield
  {title} {\bibinfo {title} {Polariton panorama},\ }\href
  {https://doi.org/doi:10.1515/nanoph-2020-0449} {\bibfield  {journal}
  {\bibinfo  {journal} {Nanophotonics}\ }\textbf {\bibinfo {volume} {10}},\
  \bibinfo {pages} {549} (\bibinfo {year} {2021})}\BibitemShut {NoStop}%
\bibitem [{\citenamefont {Lagoudakis}\ \emph {et~al.}(2010)\citenamefont
  {Lagoudakis}, \citenamefont {Pietka}, \citenamefont {Wouters}, \citenamefont
  {Andr\'e},\ and\ \citenamefont {Deveaud-Pl\'edran}}]{Lagoudakis2010}%
  \BibitemOpen
  \bibfield  {author} {\bibinfo {author} {\bibfnamefont {K.~G.}\ \bibnamefont
  {Lagoudakis}}, \bibinfo {author} {\bibfnamefont {B.}~\bibnamefont {Pietka}},
  \bibinfo {author} {\bibfnamefont {M.}~\bibnamefont {Wouters}}, \bibinfo
  {author} {\bibfnamefont {R.}~\bibnamefont {Andr\'e}},\ and\ \bibinfo {author}
  {\bibfnamefont {B.}~\bibnamefont {Deveaud-Pl\'edran}},\ }\bibfield  {title}
  {\bibinfo {title} {Coherent oscillations in an exciton-polariton josephson
  junction},\ }\href {https://doi.org/10.1103/PhysRevLett.105.120403}
  {\bibfield  {journal} {\bibinfo  {journal} {Phys. Rev. Lett.}\ }\textbf
  {\bibinfo {volume} {105}},\ \bibinfo {pages} {120403} (\bibinfo {year}
  {2010})}\BibitemShut {NoStop}%
\bibitem [{\citenamefont {Abbarchi}\ \emph {et~al.}(2013)\citenamefont
  {Abbarchi}, \citenamefont {Amo}, \citenamefont {Sala}, \citenamefont
  {Solnyshkov}, \citenamefont {Flayac}, \citenamefont {Ferrier}, \citenamefont
  {Sagnes}, \citenamefont {Galopin}, \citenamefont {Lema{\^i}tre},
  \citenamefont {Malpuech},\ and\ \citenamefont {Bloch}}]{Abbarchi2013}%
  \BibitemOpen
  \bibfield  {author} {\bibinfo {author} {\bibfnamefont {M.}~\bibnamefont
  {Abbarchi}}, \bibinfo {author} {\bibfnamefont {A.}~\bibnamefont {Amo}},
  \bibinfo {author} {\bibfnamefont {V.~G.}\ \bibnamefont {Sala}}, \bibinfo
  {author} {\bibfnamefont {D.~D.}\ \bibnamefont {Solnyshkov}}, \bibinfo
  {author} {\bibfnamefont {H.}~\bibnamefont {Flayac}}, \bibinfo {author}
  {\bibfnamefont {L.}~\bibnamefont {Ferrier}}, \bibinfo {author} {\bibfnamefont
  {I.}~\bibnamefont {Sagnes}}, \bibinfo {author} {\bibfnamefont
  {E.}~\bibnamefont {Galopin}}, \bibinfo {author} {\bibfnamefont
  {A.}~\bibnamefont {Lema{\^i}tre}}, \bibinfo {author} {\bibfnamefont
  {G.}~\bibnamefont {Malpuech}},\ and\ \bibinfo {author} {\bibfnamefont
  {J.}~\bibnamefont {Bloch}},\ }\bibfield  {title} {\bibinfo {title}
  {Macroscopic quantum self-trapping and josephson oscillations of exciton
  polaritons},\ }\href {https://doi.org/10.1038/nphys2609} {\bibfield
  {journal} {\bibinfo  {journal} {Nat. Phys.}\ }\textbf {\bibinfo {volume}
  {9}},\ \bibinfo {pages} {275} (\bibinfo {year} {2013})}\BibitemShut {NoStop}%
\bibitem [{\citenamefont {Deng}\ \emph {et~al.}(2002)\citenamefont {Deng},
  \citenamefont {Weihs}, \citenamefont {Santori}, \citenamefont {Bloch},\ and\
  \citenamefont {Yamamoto}}]{Deng2002}%
  \BibitemOpen
  \bibfield  {author} {\bibinfo {author} {\bibfnamefont {H.}~\bibnamefont
  {Deng}}, \bibinfo {author} {\bibfnamefont {G.}~\bibnamefont {Weihs}},
  \bibinfo {author} {\bibfnamefont {C.}~\bibnamefont {Santori}}, \bibinfo
  {author} {\bibfnamefont {J.}~\bibnamefont {Bloch}},\ and\ \bibinfo {author}
  {\bibfnamefont {Y.}~\bibnamefont {Yamamoto}},\ }\bibfield  {title} {\bibinfo
  {title} {Condensation of semiconductor microcavity exciton polaritons},\
  }\href {https://doi.org/10.1126/science.1074464} {\bibfield  {journal}
  {\bibinfo  {journal} {Science}\ }\textbf {\bibinfo {volume} {298}},\ \bibinfo
  {pages} {199} (\bibinfo {year} {2002})}\BibitemShut {NoStop}%
\bibitem [{\citenamefont {Kasprzak}\ \emph {et~al.}(2006)\citenamefont
  {Kasprzak}, \citenamefont {Richard}, \citenamefont {Kundermann},
  \citenamefont {Baas}, \citenamefont {Jeambrun}, \citenamefont {Keeling},
  \citenamefont {Marchetti}, \citenamefont {Szyma{\'{n}}ska}, \citenamefont
  {Andr{\'e}}, \citenamefont {Staehli}, \citenamefont {Savona}, \citenamefont
  {Littlewood}, \citenamefont {Deveaud},\ and\ \citenamefont
  {Dang}}]{Kasprzak2006}%
  \BibitemOpen
  \bibfield  {author} {\bibinfo {author} {\bibfnamefont {J.}~\bibnamefont
  {Kasprzak}}, \bibinfo {author} {\bibfnamefont {M.}~\bibnamefont {Richard}},
  \bibinfo {author} {\bibfnamefont {S.}~\bibnamefont {Kundermann}}, \bibinfo
  {author} {\bibfnamefont {A.}~\bibnamefont {Baas}}, \bibinfo {author}
  {\bibfnamefont {P.}~\bibnamefont {Jeambrun}}, \bibinfo {author}
  {\bibfnamefont {J.~M.~J.}\ \bibnamefont {Keeling}}, \bibinfo {author}
  {\bibfnamefont {F.~M.}\ \bibnamefont {Marchetti}}, \bibinfo {author}
  {\bibfnamefont {M.~H.}\ \bibnamefont {Szyma{\'{n}}ska}}, \bibinfo {author}
  {\bibfnamefont {R.}~\bibnamefont {Andr{\'e}}}, \bibinfo {author}
  {\bibfnamefont {J.~L.}\ \bibnamefont {Staehli}}, \bibinfo {author}
  {\bibfnamefont {V.}~\bibnamefont {Savona}}, \bibinfo {author} {\bibfnamefont
  {P.~B.}\ \bibnamefont {Littlewood}}, \bibinfo {author} {\bibfnamefont
  {B.}~\bibnamefont {Deveaud}},\ and\ \bibinfo {author} {\bibfnamefont {L.~S.}\
  \bibnamefont {Dang}},\ }\bibfield  {title} {\bibinfo {title} {Bose--einstein
  condensation of exciton polaritons},\ }\href
  {https://doi.org/10.1038/nature05131} {\bibfield  {journal} {\bibinfo
  {journal} {Nature}\ }\textbf {\bibinfo {volume} {443}},\ \bibinfo {pages}
  {409} (\bibinfo {year} {2006})}\BibitemShut {NoStop}%
\bibitem [{\citenamefont {Zhang}\ \emph {et~al.}(2014)\citenamefont {Zhang},
  \citenamefont {Wang}, \citenamefont {Brodbeck}, \citenamefont {Schneider},
  \citenamefont {Kamp}, \citenamefont {H{\"o}fling},\ and\ \citenamefont
  {Deng}}]{Zhang2014}%
  \BibitemOpen
  \bibfield  {author} {\bibinfo {author} {\bibfnamefont {B.}~\bibnamefont
  {Zhang}}, \bibinfo {author} {\bibfnamefont {Z.}~\bibnamefont {Wang}},
  \bibinfo {author} {\bibfnamefont {S.}~\bibnamefont {Brodbeck}}, \bibinfo
  {author} {\bibfnamefont {C.}~\bibnamefont {Schneider}}, \bibinfo {author}
  {\bibfnamefont {M.}~\bibnamefont {Kamp}}, \bibinfo {author} {\bibfnamefont
  {S.}~\bibnamefont {H{\"o}fling}},\ and\ \bibinfo {author} {\bibfnamefont
  {H.}~\bibnamefont {Deng}},\ }\bibfield  {title} {\bibinfo {title}
  {Zero-dimensional polariton laser in a subwavelength grating-based vertical
  microcavity},\ }\href {https://doi.org/10.1038/lsa.2014.16} {\bibfield
  {journal} {\bibinfo  {journal} {Light: Sci. Appl.}\ }\textbf {\bibinfo
  {volume} {3}},\ \bibinfo {pages} {e135} (\bibinfo {year} {2014})}\BibitemShut
  {NoStop}%
\bibitem [{\citenamefont {Zhang}\ \emph {et~al.}(2015)\citenamefont {Zhang},
  \citenamefont {Brodbeck}, \citenamefont {Wang}, \citenamefont {Kamp},
  \citenamefont {Schneider}, \citenamefont {Höfling},\ and\ \citenamefont
  {Deng}}]{Zhang2015}%
  \BibitemOpen
  \bibfield  {author} {\bibinfo {author} {\bibfnamefont {B.}~\bibnamefont
  {Zhang}}, \bibinfo {author} {\bibfnamefont {S.}~\bibnamefont {Brodbeck}},
  \bibinfo {author} {\bibfnamefont {Z.}~\bibnamefont {Wang}}, \bibinfo {author}
  {\bibfnamefont {M.}~\bibnamefont {Kamp}}, \bibinfo {author} {\bibfnamefont
  {C.}~\bibnamefont {Schneider}}, \bibinfo {author} {\bibfnamefont
  {S.}~\bibnamefont {Höfling}},\ and\ \bibinfo {author} {\bibfnamefont
  {H.}~\bibnamefont {Deng}},\ }\bibfield  {title} {\bibinfo {title} {Coupling
  polariton quantum boxes in sub-wavelength grating microcavities},\ }\href
  {https://doi.org/10.1063/1.4907606} {\bibfield  {journal} {\bibinfo
  {journal} {Appl. Phys. Lett.}\ }\textbf {\bibinfo {volume} {106}},\ \bibinfo
  {pages} {051104} (\bibinfo {year} {2015})}\BibitemShut {NoStop}%
\bibitem [{\citenamefont {Li}\ \emph {et~al.}(2023)\citenamefont {Li},
  \citenamefont {Lomsadze}, \citenamefont {Moody}, \citenamefont {Smallwood},\
  and\ \citenamefont {Cundiff}}]{MDCS_Book}%
  \BibitemOpen
  \bibfield  {author} {\bibinfo {author} {\bibfnamefont {H.}~\bibnamefont
  {Li}}, \bibinfo {author} {\bibfnamefont {B.}~\bibnamefont {Lomsadze}},
  \bibinfo {author} {\bibfnamefont {G.}~\bibnamefont {Moody}}, \bibinfo
  {author} {\bibfnamefont {C.}~\bibnamefont {Smallwood}},\ and\ \bibinfo
  {author} {\bibfnamefont {S.}~\bibnamefont {Cundiff}},\ }\href@noop {} {\emph
  {\bibinfo {title} {Optical Multidimensional Coherent Spectroscopy}}}\
  (\bibinfo  {publisher} {Oxford University Press},\ \bibinfo {address}
  {Oxford},\ \bibinfo {year} {2023})\BibitemShut {NoStop}%
\bibitem [{\citenamefont {Takemura}\ \emph {et~al.}(2015)\citenamefont
  {Takemura}, \citenamefont {Trebaol}, \citenamefont {Anderson}, \citenamefont
  {Kohnle}, \citenamefont {L\'eger}, \citenamefont {Oberli}, \citenamefont
  {Portella-Oberli},\ and\ \citenamefont {Deveaud}}]{Takemura2015}%
  \BibitemOpen
  \bibfield  {author} {\bibinfo {author} {\bibfnamefont {N.}~\bibnamefont
  {Takemura}}, \bibinfo {author} {\bibfnamefont {S.}~\bibnamefont {Trebaol}},
  \bibinfo {author} {\bibfnamefont {M.~D.}\ \bibnamefont {Anderson}}, \bibinfo
  {author} {\bibfnamefont {V.}~\bibnamefont {Kohnle}}, \bibinfo {author}
  {\bibfnamefont {Y.}~\bibnamefont {L\'eger}}, \bibinfo {author} {\bibfnamefont
  {D.~Y.}\ \bibnamefont {Oberli}}, \bibinfo {author} {\bibfnamefont {M.~T.}\
  \bibnamefont {Portella-Oberli}},\ and\ \bibinfo {author} {\bibfnamefont
  {B.}~\bibnamefont {Deveaud}},\ }\bibfield  {title} {\bibinfo {title}
  {Two-dimensional fourier transform spectroscopy of exciton-polaritons and
  their interactions},\ }\href {https://doi.org/10.1103/PhysRevB.92.125415}
  {\bibfield  {journal} {\bibinfo  {journal} {Phys. Rev. B}\ }\textbf {\bibinfo
  {volume} {92}},\ \bibinfo {pages} {125415} (\bibinfo {year}
  {2015})}\BibitemShut {NoStop}%
\bibitem [{\citenamefont {Xiang}\ and\ \citenamefont
  {Xiong}(2024)}]{Xiang2024}%
  \BibitemOpen
  \bibfield  {author} {\bibinfo {author} {\bibfnamefont {B.}~\bibnamefont
  {Xiang}}\ and\ \bibinfo {author} {\bibfnamefont {W.}~\bibnamefont {Xiong}},\
  }\bibfield  {title} {\bibinfo {title} {Molecular polaritons for chemistry,
  photonics and quantum technologies},\ }\href
  {https://doi.org/10.1021/acs.chemrev.3c00662} {\bibfield  {journal} {\bibinfo
   {journal} {Chem. Rev.}\ }\textbf {\bibinfo {volume} {124}},\ \bibinfo
  {pages} {2512} (\bibinfo {year} {2024})}\BibitemShut {NoStop}%
\bibitem [{\citenamefont {Wen}\ \emph {et~al.}(2013)\citenamefont {Wen},
  \citenamefont {Christmann}, \citenamefont {Baumberg},\ and\ \citenamefont
  {Nelson}}]{Wen2013}%
  \BibitemOpen
  \bibfield  {author} {\bibinfo {author} {\bibfnamefont {P.}~\bibnamefont
  {Wen}}, \bibinfo {author} {\bibfnamefont {G.}~\bibnamefont {Christmann}},
  \bibinfo {author} {\bibfnamefont {J.~J.}\ \bibnamefont {Baumberg}},\ and\
  \bibinfo {author} {\bibfnamefont {K.~A.}\ \bibnamefont {Nelson}},\ }\bibfield
   {title} {\bibinfo {title} {Influence of multi-exciton correlations on
  nonlinear polariton dynamics in semiconductor microcavities},\ }\href
  {https://doi.org/10.1088/1367-2630/15/2/025005} {\bibfield  {journal}
  {\bibinfo  {journal} {New J. Phys.}\ }\textbf {\bibinfo {volume} {15}},\
  \bibinfo {pages} {025005} (\bibinfo {year} {2013})}\BibitemShut {NoStop}%
\bibitem [{\citenamefont {Autry}\ \emph {et~al.}(2020)\citenamefont {Autry},
  \citenamefont {Nardin}, \citenamefont {Smallwood}, \citenamefont {Silverman},
  \citenamefont {Bajoni}, \citenamefont {Lema\^{\i}tre}, \citenamefont
  {Bouchoule}, \citenamefont {Bloch},\ and\ \citenamefont
  {Cundiff}}]{Autry2020}%
  \BibitemOpen
  \bibfield  {author} {\bibinfo {author} {\bibfnamefont {T.~M.}\ \bibnamefont
  {Autry}}, \bibinfo {author} {\bibfnamefont {G.}~\bibnamefont {Nardin}},
  \bibinfo {author} {\bibfnamefont {C.~L.}\ \bibnamefont {Smallwood}}, \bibinfo
  {author} {\bibfnamefont {K.}~\bibnamefont {Silverman}}, \bibinfo {author}
  {\bibfnamefont {D.}~\bibnamefont {Bajoni}}, \bibinfo {author} {\bibfnamefont
  {A.}~\bibnamefont {Lema\^{\i}tre}}, \bibinfo {author} {\bibfnamefont
  {S.}~\bibnamefont {Bouchoule}}, \bibinfo {author} {\bibfnamefont
  {J.}~\bibnamefont {Bloch}},\ and\ \bibinfo {author} {\bibfnamefont
  {S.}~\bibnamefont {Cundiff}},\ }\bibfield  {title} {\bibinfo {title}
  {Excitation ladder of cavity polaritons},\ }\href
  {https://doi.org/10.1103/PhysRevLett.125.067403} {\bibfield  {journal}
  {\bibinfo  {journal} {Phys. Rev. Lett.}\ }\textbf {\bibinfo {volume} {125}},\
  \bibinfo {pages} {067403} (\bibinfo {year} {2020})}\BibitemShut {NoStop}%
\bibitem [{\citenamefont {Nardin}\ \emph {et~al.}(2013)\citenamefont {Nardin},
  \citenamefont {Autry}, \citenamefont {Silverman},\ and\ \citenamefont
  {Cundiff}}]{Nardin2013}%
  \BibitemOpen
  \bibfield  {author} {\bibinfo {author} {\bibfnamefont {G.}~\bibnamefont
  {Nardin}}, \bibinfo {author} {\bibfnamefont {T.~M.}\ \bibnamefont {Autry}},
  \bibinfo {author} {\bibfnamefont {K.~L.}\ \bibnamefont {Silverman}},\ and\
  \bibinfo {author} {\bibfnamefont {S.~T.}\ \bibnamefont {Cundiff}},\
  }\bibfield  {title} {\bibinfo {title} {Multidimensional coherent photocurrent
  spectroscopy of a semiconductor nanostructure},\ }\href
  {https://doi.org/10.1364/OE.21.028617} {\bibfield  {journal} {\bibinfo
  {journal} {Opt. Express}\ }\textbf {\bibinfo {volume} {21}},\ \bibinfo
  {pages} {28617} (\bibinfo {year} {2013})}\BibitemShut {NoStop}%
\bibitem [{\citenamefont {Martin}\ \emph {et~al.}(2020)\citenamefont {Martin},
  \citenamefont {Horng}, \citenamefont {Ruth}, \citenamefont {Paik},
  \citenamefont {Wentzel}, \citenamefont {Deng},\ and\ \citenamefont
  {Cundiff}}]{Martin2020}%
  \BibitemOpen
  \bibfield  {author} {\bibinfo {author} {\bibfnamefont {E.~W.}\ \bibnamefont
  {Martin}}, \bibinfo {author} {\bibfnamefont {J.}~\bibnamefont {Horng}},
  \bibinfo {author} {\bibfnamefont {H.~G.}\ \bibnamefont {Ruth}}, \bibinfo
  {author} {\bibfnamefont {E.}~\bibnamefont {Paik}}, \bibinfo {author}
  {\bibfnamefont {M.-H.}\ \bibnamefont {Wentzel}}, \bibinfo {author}
  {\bibfnamefont {H.}~\bibnamefont {Deng}},\ and\ \bibinfo {author}
  {\bibfnamefont {S.~T.}\ \bibnamefont {Cundiff}},\ }\bibfield  {title}
  {\bibinfo {title} {Encapsulation narrows and preserves the excitonic
  homogeneous linewidth of exfoliated monolayer
  ${\mathrm{mo}\mathrm{se}}_{2}$},\ }\href
  {https://doi.org/10.1103/PhysRevApplied.14.021002} {\bibfield  {journal}
  {\bibinfo  {journal} {Phys. Rev. Appl.}\ }\textbf {\bibinfo {volume} {14}},\
  \bibinfo {pages} {021002} (\bibinfo {year} {2020})}\BibitemShut {NoStop}%
\bibitem [{\citenamefont {Liu}\ \emph {et~al.}(2024)\citenamefont {Liu},
  \citenamefont {Pavi{\'{c}}evi{\'{c}}}, \citenamefont {Michael}, \citenamefont
  {Salvador}, \citenamefont {Dolgirev}, \citenamefont {Fechner}, \citenamefont
  {Disa}, \citenamefont {M.~Lozano}, \citenamefont {Li}, \citenamefont {Gu},
  \citenamefont {Demler},\ and\ \citenamefont {Cavalleri}}]{Liu_2024_NatPhys}%
  \BibitemOpen
  \bibfield  {author} {\bibinfo {author} {\bibfnamefont {A.}~\bibnamefont
  {Liu}}, \bibinfo {author} {\bibfnamefont {D.}~\bibnamefont
  {Pavi{\'{c}}evi{\'{c}}}}, \bibinfo {author} {\bibfnamefont {M.~H.}\
  \bibnamefont {Michael}}, \bibinfo {author} {\bibfnamefont {A.~G.}\
  \bibnamefont {Salvador}}, \bibinfo {author} {\bibfnamefont {P.~E.}\
  \bibnamefont {Dolgirev}}, \bibinfo {author} {\bibfnamefont {M.}~\bibnamefont
  {Fechner}}, \bibinfo {author} {\bibfnamefont {A.~S.}\ \bibnamefont {Disa}},
  \bibinfo {author} {\bibfnamefont {P.}~\bibnamefont {M.~Lozano}}, \bibinfo
  {author} {\bibfnamefont {Q.}~\bibnamefont {Li}}, \bibinfo {author}
  {\bibfnamefont {G.~D.}\ \bibnamefont {Gu}}, \bibinfo {author} {\bibfnamefont
  {E.}~\bibnamefont {Demler}},\ and\ \bibinfo {author} {\bibfnamefont
  {A.}~\bibnamefont {Cavalleri}},\ }\bibfield  {title} {\bibinfo {title}
  {Probing inhomogeneous cuprate superconductivity by terahertz josephson echo
  spectroscopy},\ }\href {https://doi.org/10.1038/s41567-024-02643-5}
  {\bibfield  {journal} {\bibinfo  {journal} {Nat. Phys.}\ }\textbf {\bibinfo
  {volume} {20}},\ \bibinfo {pages} {1751} (\bibinfo {year}
  {2024})}\BibitemShut {NoStop}%
\end{thebibliography}%

\end{document}